# Magnon transport and thermoelectric effects in ultrathin $Tm_3Fe_5O_{12}$/Pt nonlocal devices

Jialiang Gao[1,†], Charles-Henri Lambert[1], Richard Schlitz[1], Manfred Fiebig[2], Pietro Gambardella[1], and Saül Vélez[1,2,3,*]

[1]Laboratory for Magnetism and Interface Physics, Department of Materials, ETH Zürich, 8093 Zürich, Switzerland

[2]Laboratory for Multifunctional Ferroic Materials, Department of Materials, ETH Zürich, 8093 Zürich, Switzerland

[3]Spintronics and Nanodevices Laboratory, Condensed Matter Physics Center (IFIMAC), Instituto Nicolás Cabrera, and Departamento de Física de la Materia Condensada, Universidad Autónoma de Madrid, E-28049 Madrid, Spain

[†]Present address: Photonics Laboratory, Department of Information Technology and Electrical Engineering, ETH Zürich, 8093 Zürich, Switzerland

*saul.velez@uam.es

The possibility of electrically exciting and detecting magnon currents in magnetic insulators has opened exciting perspectives for transporting spin information in electronic devices. However, the role of the magnetic field and the nonlocal thermal gradients on the magnon transport remain unclear. Here, by performing nonlocal harmonic voltage measurements, we investigate magnon transport in perpendicularly magnetized ultrathin $Tm_3Fe_5O_{12}$ (TmIG) films coupled to Pt electrodes. We show that the first harmonic nonlocal voltage captures spin-driven magnon transport in TmIG, as expected, and the second harmonic is dominated by thermoelectric voltages driven by current-induced thermal gradients at the detector. The magnon diffusion length in TmIG is found to be $\lambda_m \sim 0.3$ μm at 0.5 T and gradually decays to $\lambda_m \sim 0.2$ μm at 0.8 T, which we attribute to the suppression of the magnon relaxation time due to the increase of the Gilbert damping with field. By performing current, magnetic field, and distance dependent nonlocal and local measurements we demonstrate that the second harmonic nonlocal voltage exhibits five thermoelectric contributions, which originate from the nonlocal spin Seebeck effect and the ordinary, planar, spin, and anomalous Nernst effects. Our work provides a guide on how to disentangle magnon signals from diverse thermoelectric voltages of spin and magnetic origin in nonlocal magnon devices, and establish the scaling laws of the thermoelectric voltages in metal/insulator bilayers.

## I. INTRODUCTION

Magnetic insulators (MI) coupled to heavy metal (HM) layers are an ideal platform for the development of novel, yet highly-efficient, spintronic devices [1] [2] [3]. For instance, the low magnetic moment and low Gilbert damping of thin film ferrimagnetic garnets enable to stabilize chiral domain walls and electrically drive them with mobilities exceeding those attained with conventional ferromagnets [4] [5] [6]. Moreover, unlike in metals, magnon currents in MIs can propagate spin information over several micrometers, which can be electrically excited and detected by exploiting charge-to-spin conversion phenomena in HMs [7] [8] [9].

Magnon current injection and diffusion is typically investigated in MI/HM nonlocal device structures as the one shown in Fig. 1a. The charge current injected in one HM stripe can drive magnon currents in the MI via both spin and heat transfer across the MI/HM interface (the charge current generates both a spin current via the spin Hall effect (SHE) [10] and a thermal gradient via Joule heating; from now on, we will refer to these mechanisms as "*electrical*" and "*thermal*", respectively), which are subsequently detected in a second nonlocal HM stripe by exploiting the reciprocal processes, i.e, the magnon-to-spin conversion at the MI/HM interface and the spin-to-charge conversion in the HM [8]. Magnon transport has been proven to be very efficient in low-damping materials such as $Y_3Fe_5O_{12}$ (YIG) [7] [8] [11] [12] [13] [14] and $\alpha-Fe_2O_3$ [9] [15] [16]. However, several questions and challenges remain open. For instance, while earlier reports indicated that the magnon currents driven by electrical and thermal means should have the same propagation length [8] [17], more recent works suggest they are different [14]. In addition, the generation of thermal magnon currents also result in nonlocal thermal gradients, which can lead to voltages of thermoelectric origin at the detector [18] and thus complicate the identification of the diffusive magnon currents. Besides, the influence of the external magnetic field on the transport characteristics of the magnon currents is not fully clear [14] [19] [20]. Yet, detailed magnon transport experiments have only been reported in a limited number of materials, typically involving relatively thick layers (>100 nm) of YIG. Therefore, it is essential that new materials with complementary properties are investigated. Furthermore, it is necessary to establish the scaling laws of the thermoelectric voltages in nonlocal devices to unambiguously disentangle diffusive magnon signals from spurious effects.

In this work, we report magnon transport experiments in sputter-grown perpendicularly magnetized ultrathin TmIG layers. By performing harmonic transport measurements in patterned nonlocal Pt devices, we can separate the signals arising from electrically generated magnon currents from those of thermal origin. The magnon diffusion length is determined by analyzing the exponential decay of the first harmonic signal for devices having different injector-detector distances. Moreover, we show that the external field leads to a reduction of the magnon diffusion length in TmIG due to the suppression of the magnon relaxation time with field. We explain the field dependence based on the Gilbert damping phenomenology and derive simple relations that describes the magnon transport for moderate field perturbations. We also show that the thermally-induced signals are captured in the second harmonic nonlocal response and are, however, dominated by the nonlocal thermal gradients induced nearby the detector rather than by diffusive magnon currents propagating from the injector. We identify five different thermoelectric contributions, namely, a nonlocal spin Seebeck effect (SSE) [18], an ordinary Nernst effect (ONE), a planar Nernst effect (PNE) [21], a spin Nernst effect (SNE) [22], and a thermoelectric anomalous Hall-like voltage originating from the anomalous Nernst effect (ANE) [23]. Finally, we show that by performing angular, magnetic field, current, and distance dependence experiments it is possible to discern between thermal magnon currents and nonlocal SSE voltages [18], as well as to distinguish proximity-induced signals (such as the PNE and the ANE) from those having same symmetry but spin origin (i.e., the SNE and the Thermal Spin Drag (TSD) [24], respectively) in nonlocal devices.

## II. EXPERIMENTAL DETAILS

*Thin films growth and characterization*. The films were prepared by in situ sputter growth of TmIG and Pt. A 15-nm-thick TmIG layer was rf-sputtered on Sc-substituted gadolinium gallium garnet single crystal substrates ($Gd_3Sc_2Ga_3O_{12}$; GSGG) with a base pressure of $5 \times 10^{-8}$ Torr. The growth temperature and the Ar pressure for the TmIG were 800 ℃ and 0.2 Pa, respectively. After the deposition, the samples were annealed for 30 min at the deposition temperature and finally cooled back to room temperature in vacuum. In order to improve the TmIG/Pt interface quality, 4 nm Pt was dc-sputtered at room temperature in the same chamber without breaking the vacuum. The Ar pressure during the Pt deposition was 0.4 Pa. The crystalline structure and topography of the layers were characterized by x-ray diffraction and atomic force microscopy measurements, the later revealing a root-mean-square roughness below 1 nm over a 5 μm² surface area (see Appendix A for more details). The saturation magnetization of the 15-nm-TmIG film investigated in this work was determined to be $M_s \sim 100 \text{ kA m}^{-1}$, which agrees with the reduction of $M_s$ in films relative to the bulk value [4] [25] [26] [27]. The anisotropy was characterized by spin Hall magnetoresistance (SMR) measurements in Hall bar devices [4] [27] [28] [29], demonstrating that the film exhibits robust perpendicular magnetic anisotropy with an anisotropy field ~400 mT and a coercive field $H_c \sim 30$ mT (see Appendix B).

*Device fabrication*. Hall bars and nonlocal devices having different distances between the Pt electrodes (center-to-center distances ranging from $d = 1.0$ to $4.1$ µm and fixed width $w \sim 0.5$ µm) were patterned on the same film by high-resolution electron beam lithography and Ar plasma etching. To that end, a 3-nm-thick Ti hard mask was patterned on the films by writing the structures on a polymethyl methacrylate layer, followed by the e-beam evaporation of Ti and lift off. The final devices were produced by etching the uncovered Pt regions in low-power Ar$^+$ plasma, leaving the TmIG film unetched. Fig. 1(a) shows an optical image of a representative nonlocal device with $d = 2.5$ µm. The long and short stripes have a length $L = 185$ and $125$ µm, respectively. Fig. 8(a) in Appendix B shows an optical image of a Hall bar device, which was used to perform local magnetotransport measurements. Note that the Ti layer thickness was optimized to result in a ∼1-nm-thick non-conducting TiO$_x$ capping layer on the Pt devices after etching.

*Harmonic transport measurements*. The transport measurements were performed in a room temperature magnetotransport setup that allowed us to rotate the sample by 360° along any of the three main rotation axes of the devices. Harmonic transport measurements were performed by demodulating the detected voltage at integer multiples of the frequency of the injected alternating current (ac). This was realized by employing a sinusoidal current source synchronized with a multi-channel high-frequency PXI acquisition card module operating in continuous mode. A fast Fourier transformation of the voltage signals allows extracting the amplitude and phase of the different harmonics. This measurement scheme enables us to capture the voltage response of the device in multiple channels, which we employed to simultaneously record the local and nonlocal response of different Pt electrodes in nonlocal devices, as well as the longitudinal and transverse voltages in Hall bar structures. In the nonlocal measurements, the frequency of the injected current was fixed to 3 Hz to avoid spurious inductive signals. No current leakage through TmIG was detected in the current range employed in the experiments.

### III. RESULTS AND DISCUSSION

#### A. Electrically- and thermally-induced nonlocal signals

The electrically- and thermally-induced signals can be decoupled by performing harmonic transport measurements. The first harmonic nonlocal response captures the signal arising from the electrical excitation of magnon currents (i.e., due to the charge-to-spin and the spin-to-magnon conversion), whereas the second harmonic proves the thermally-driven magnons as well as thermoelectric signals arising from the thermal gradients induced at the detector [8]. An optical image of a representative nonlocal device investigated in this work is presented in Fig. 1(a).

Due to the symmetries of the effects involved in the excitation and detection of magnon currents, the nonlocal signals have characteristic angular dependences. Figs. 1(b) and 1(c) show representatives first and second harmonic nonlocal voltage measurements performed with a magnetic field **B** rotating in the plane of the film. The strength of the magnetic field is selected to saturate the magnetization **M** of TmIG along **B**. The first harmonic response shows a $\sin^2\alpha$ dependence [$\alpha$ is the angle between the injected current and **M** in the plane of the film; see Fig. 1(a)], reflecting the symmetry of the SHE and the inverse SHE at the injector and detector electrodes [8] [14] [17], respectively. Concretely, the spin-to-magnon excitation is maximal (zero) for **M** collinear (perpendicular) to the spin accumulation **μ**$_s$ at the TmIG/Pt interface, which is in the plane and perpendicular to the stripe, giving a $\sin\alpha$ dependence for magnon excitation at the injector. The second $\sin\alpha$ arises from the magnon to spin conversion at the detector, as the detection of the inverse SHE voltage is proportional to the alignment of **M** in the direction perpendicular to the stripe. We thus confirm that the first harmonic signal in our TmIG/Pt nonlocal devices captures electrically driven diffusive magnons in TmIG.

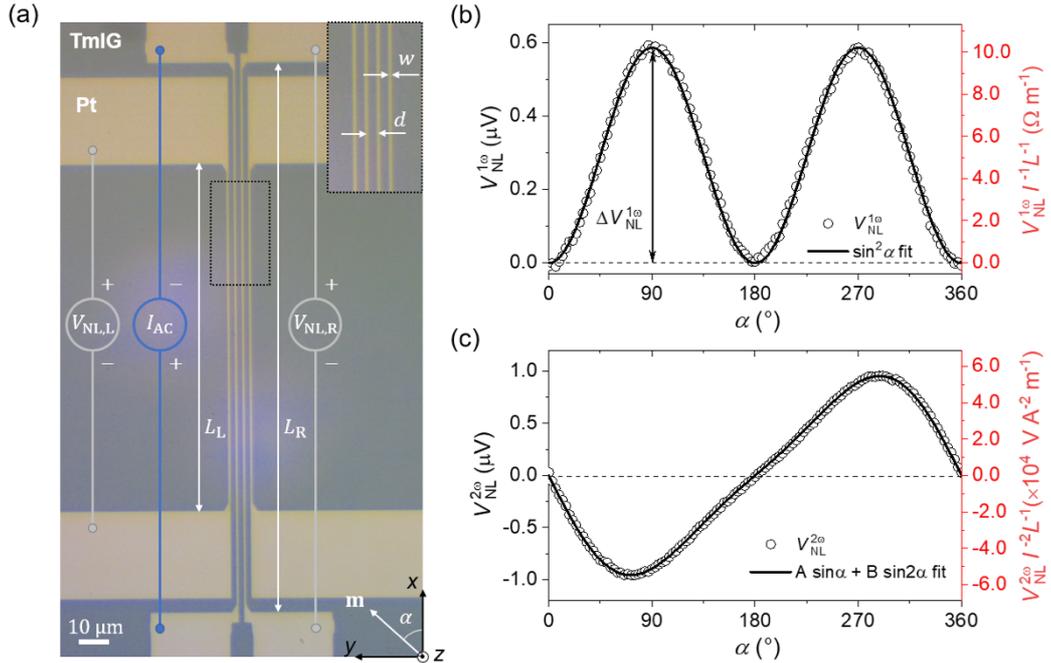

FIG. 1. (a) Optical image of a representative nonlocal device. The coordinate system, the rotation angle $\alpha$, and the electric wiring employed in nonlocal measurements are indicated. The voltage in the current line is also measured as reference (not indicated for simplicity). The inset shows a zoom of the dashed area. $w = 0.5$ μm, $d = 2.5$ μm, $L_\text{R} = 185$ μm, and $L_\text{L} = 125$ μm. (b), (c) Angular dependent measurements of the first and second nonlocal voltage response $V_\text{NL}^{1\omega}$ and $V_\text{NL}^{2\omega}$ captured in the electrode located at the right-hand side of the current line for a device with $d = 1.2$ μm. The measurements are performed with an ac current excitation of amplitude $I = 310$ μA and external field $B = 500$ mT. The solid lines are fits to the experimental data. The right Y-scale in (b) and (c) shows the $V_{\text{NL}}^{1\omega}$ and $V_{\text{NL}}^{2\omega}$ voltages normalized by the current (linear and quadratic, respectively) and length of the detector.

The second harmonic signal has two contributions, one exhibiting a $\sin\alpha$ dependence and the other $\sin 2\alpha$ [see Fig. 1(c)]. The $\sin\alpha$ dependence is consistent with the detection of thermally driven magnon currents (which diffuse away from the injector region to the detector electrode and can be parametrized by the magnon chemical potential), as only the inverse SHE at the detector is involved [8] [14] [30]. However, the same symmetry is also observed for thermoelectric signals originated by vertical thermal gradients (i.e., perpendicular to the film plane) present at the detector region [18]. For thin MI layers, the vertical thermal gradient causes a magnon accumulation at the MI/substrate interface underneath the detector, resulting in a vertical diffusion and detection of magnons at the HM electrode. Because of its physical origin, we identify this thermoelectric signal as *nonlocal SSE.* Although both mechanisms result in signals having identical angular dependences, their origin stem from physical parameters having distinct decay lengths, i.e., the magnon chemical potential and the thermal diffusion, which decay exponentially and geometrically, respectively. Consequently, the thermal magnon currents can be distinguished from the nonlocal SSE by analyzing the decay of the signal with the injector-detector distance [14] [18].

The $\sin 2\alpha$ contribution is consistent with a thermoelectric signal arising from the in-plane thermal gradient present at the detector electrode [22] [24]. However, we note that in YIG/Pt nonlocal structures, the $\sin 2\alpha$ contribution was typically found to be negligible compared to the $\sin\alpha$ signal [8] [14] [22]. The origin of the different contributions to the second harmonic thermal signal in TmIG/Pt is analyzed in detail in Section C. The distance and magnetic field dependences of the first harmonic, which include the analysis of the magnon diffusion length in TmIG, are presented in Section B.

**B. Electrically-driven magnon currents**

**B.1. Determination of the magnon diffusion length**

The magnon diffusion length can be extracted by measuring the decay of the first harmonic nonlocal signal with the injector-detector distance $d$. Fig. 2(a) shows the amplitude of the first harmonic signal $\Delta V_{\text{NL}}^{1\omega}$ (see Fig. 1(b)) measured in devices with different $d$ and constant magnetic field. As expected for diffusive magnons, in the relaxation regime, i.e., for distances similar or longer than the magnon diffusion length $\lambda_{\text{m}}$, and for film thicknesses $t < \lambda_{\text{m}}$, the nonlocal voltage decays with $d$ as [8] [17]

$$\Delta V_{\text{NL}}^{1\omega} = IL\frac{C}{\lambda_{\text{m}}}\exp\left(-\frac{d}{\lambda_{\text{m}}}\right), \tag{1}$$

where $C$ is a distance independent prefactor that depends on different parameters of the MI as well as the spin transmission efficiency across the HM/MI interface. As demonstrated in Fig. 2(b), the signal decays exponentially with $d$ for $d > 1.0$ µm, from which we determine the magnon diffusion length of TmIG to be $\lambda_{\text{m}}(0.5\text{ T}) = 0.28 \pm 0.02$ µm. This value is similar to the one indirectly estimated for TmIG via spin wave spectroscopic measurements [31] but significantly lower than the magnon diffusion length measured in YIG thick films [7] [8] [11] [12] [13] [14], α–$Fe_2O_3$ [9] [15], or $NiFe_2O_4$ [32] [33]. The lower $\lambda_{\text{m}}$ of TmIG can be explained by the larger damping $\alpha$ of this material ($\sim 10^{-2}$ [34] [35]) compared to $NiFe_2O_4$ ($\sim 3.5 \cdot 10^{-3}$ [36]), α–$Fe_2O_3$ ($\sim 10^{-4}$ [9] [37]), and YIG ($\sim 10^{-4} - 10^{-5}$ even in thin films [38] [39]). However, the magnon diffusion length in YIG films was found to decrease drastically upon reducing the film thickness [40] [41], suggesting that not only the damping but also the scattering at the film surfaces may influence the magnon transport. In addition, we note that the magnon dispersion may be altered due to vertical confinement and thus influence the magnon manifold that contribute to the magnon transport. These questions, which go beyond the scope of this work, could be addressed via combined transport and spatially-resolved Brillouin light scattering experiments [42].

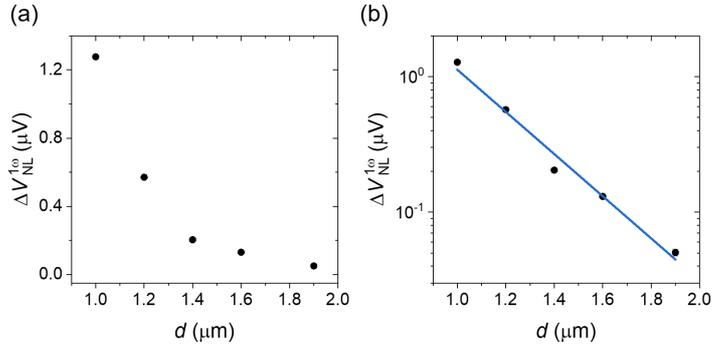

FIG. 2. (a) Amplitude of the first harmonic voltage $\Delta V_{\text{NL}}^{1\omega}$ measured for different injector-detector distances $d$. $B = 500$ mT, $L = 125$ µm, and $I = 310$ µA. The error bars are smaller than the size of the points. (b) Same data presented in (a) in logarithmic scale, showing that $\Delta V_{\text{NL}}^{1\omega}$ decays exponentially with $d$. The blue line is the fit of the data in the log scale to Eq. (1).

### B.2. Field dependence of the magnon diffusion length

The magnon diffusion length in TmIG decays with magnetic field $B$, as evidenced by a steeper decrease of $\Delta V_{\text{NL}}^{1\omega}$ with $d$ as $B$ increases (Fig. 3). A reduction of $\lambda_{\text{m}}$ with $B$ has been observed in other

MIs [9] [14] [19] [20]. However, not only the field dependence of $\lambda_m$ influences $\Delta V_{NL}^{1\omega}$, but also that of the prefactor $C(B)$ [19] [see Eq. (1)]. To investigate the role of $B$ on the magnon transport characteristics, we measured the field dependence of $\Delta V_{NL}^{1\omega}$ by recording the nonlocal voltage while sweeping $B$ in the plane of the film and perpendicular to the current line (i.e., parallel to $\boldsymbol{\mu}_s$). Fig. 3(a) shows a representative $\Delta V_{NL}^{1\omega}(B)$ curve measured in a device with $d = 1.2$ µm. For fields below $|B| \sim 400$ mT [region indicated with a grayish-colored background in Fig. 3(a)], the magnetic field gradually tilts **M** of TmIG towards the plane (see Appendix B), explaining the increase of the nonlocal signal as $B$ increases. A further increase of $|B|$, however, results in a suppression of the nonlocal signal. The same behavior with field is observed for different injector-detector distances [see Fig. 3(b), which presents $\Delta V_{NL}^{1\omega}(B)$ data for $B > 400$ mT; note that regardless of the field, the amplitude of the signal is smaller as $d$ increases].

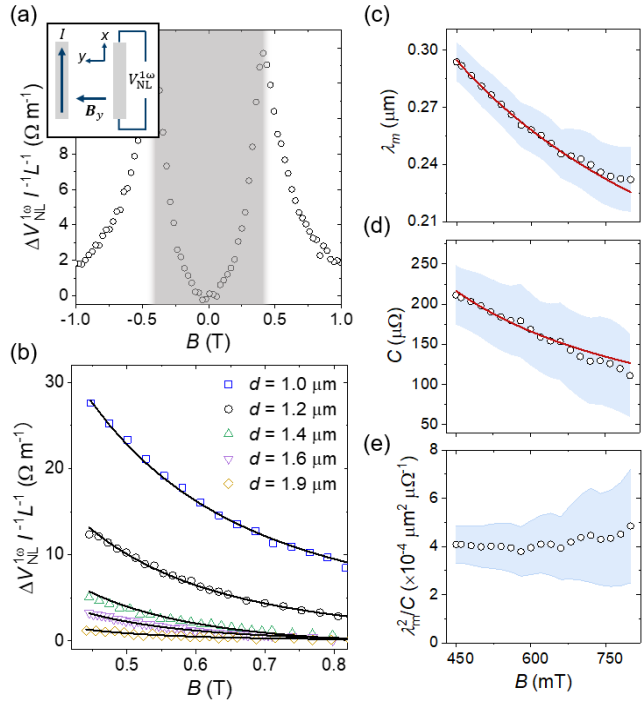

FIG. 3. (a) Field traces of $\Delta V_{NL}^{1\omega}$ measured for **B** applied in the plane of the film and perpendicular to the current line. See inset for the measurement configuration. The grey shaded area indicates the field region in which **M** of TmIG is gradually aligning with **B** (see Appendix B). $I = 310$ µA, $d = 1.2$ µm, and $L = 185$ µm. (b) $\Delta V_{NL}^{1\omega}(B)$ measured in devices with different injector-detector distances. Data shown for **M** collinear to **B** (open symbols). The injection current is $I = 310$ µA. The solid black lines are fits to the data by using Eq. (4). See text for details. (c), (d) Field dependence of $\lambda_m$ and $C$ extracted from fitting the data in (b) to Eq. (1) up to 800 mT (open symbols). Red solid lines in (c) and (d) are fits to the data points according to Eqs. (2) and (3), respectively. (e) $\lambda_m^2/C$ computed from (c) and (d). The dashed blue area in (c)-(d) and (e) represent half the standard deviation of the fitting and the corresponding error propagation.

$\lambda_\mathrm{m}(B)$ and $C(B)$ can be obtained by fitting the decay of the nonlocal signal with $d$ to Eq. (1) for different field values. Figs. 3(c) and 3(d) show $\lambda_\mathrm{m}(B)$ and $C(B)$ extracted from the data shown in Fig. 3b in the field range $450 - 800$ mT. For $B > 800$ mT, the fits are not reliable due to the low signal-to-noise ratio in the devices with larger distances. $\lambda_\mathrm{m}$ gradually reduces with $B$ from $\lambda_\mathrm{m} = 0.29 \pm 0.01$ $\mu$m at 450 mT to $\lambda_\mathrm{m} = 0.23 \pm 0.02$ $\mu$m at 800 mT. This corresponds to a relative change of $\lambda_\mathrm{m}$ of about 59% per Tesla field applied, which is about 35% larger than the one found for YIG in the same field range [19]. Although the uncertainty in the estimate of $C(B)$ is large, the analysis shows that it also reduces with $B$ (Fig. 3(d)).

To address the field dependence of $\lambda_\mathrm{m}$ and $C$, we need to consider that $\lambda_\mathrm{m} = \sqrt{D_\mathrm{m}\tau}$ and $C \propto D_\mathrm{m} n$, where $\tau$ is the magnon relaxation time, $D_\mathrm{m}$ the magnon diffusion constant, and $n$ the equilibrium magnon density [8]. Therefore, the magnetic field dependence of the magnon transport in TmIG could, in principle, be attributed to the field dependence of either $\tau$, $n$, and $D_\mathrm{m}$ [19] [43] [44] [45]. In fact, previous reports in YIG revealed that more than one transport parameter should influence the field dependence of the magnon diffusion [19], but it was not possible to identify which one was dominating, nor identify their field dependences. In the following, we will argue what field dependences are expected for $\tau$, $n$, and $D_\mathrm{m}$ in the linear regime based on simple modeling and comparison with the magnon transport data in TmIG.

The magnetic field dependence of $\tau$ can be understood in the framework of the Gilbert damping phenomenology: the increase of the magnetic field increases the frequency of the magnon modes, resulting in an enhancement of the magnon damping. In this context, it has been shown that the lifetime of the long wavelength magnons, which are the ones that dominate the magnon transport, is given by $\tau(B) \approx \tau_0/(1 + \xi B)$ [46] [47] [48], where $\tau_0$ is the magnon relaxation time at zero field, and $\xi$ represents the enhancement of the magnon damping with field. By considering that $\lambda_\mathrm{m} = \sqrt{D_\mathrm{m}\tau}$, the field dependence of $\lambda_\mathrm{m}$ can be written as

$$\lambda_\mathrm{m} \approx \frac{\lambda_{\mathrm{m},0}}{\sqrt{1+\xi B}}, \qquad (2)$$

where $\lambda_{\mathrm{m},0} = \sqrt{D_\mathrm{m}\tau_0}$ is the magnon diffusion length at zero field and $D_\mathrm{m}$ a field independent diffusion constant parameter. To find the field dependence of $C$, we computed $\frac{\lambda_\mathrm{m}^2}{C} \propto \tau/n$ and found it to be fairly constant in the field range $450 - 800$ mT (Fig. 3(e)), indicating that the magnon population decays with the magnetic field in a similar fashion as $\tau(B)$. Moreover, the slight upturn of $\tau/n$ in the range $700 - 800$ mT suggests that $n(B)$ may decay slightly faster with field than $\tau(B)$, but the large uncertainty in the estimate of $\tau/n$ prevents us to infer the exact field dependence of $n(B)$ in the high field range. Assuming $\tau/n$ to be constant with field, the field dependence of $C$ can be written as

$$C \approx \frac{C_0}{1+\xi B}, \qquad (3)$$

where $C_0 \propto n_0 D_\mathrm{m}$ is the value of $C$ at zero field with $n_0 = n(0)$. Using Eqs. (2) and (3), the field dependence of $\Delta V_\mathrm{NL}^{1\omega}$ can be expressed as

$$\Delta V_\mathrm{NL}^{1\omega}(B) \approx IL \frac{C_0}{\lambda_{\mathrm{m},0}\sqrt{1+\xi B}} \exp\left(-\frac{d\sqrt{1+\xi B}}{\lambda_{\mathrm{m},0}}\right). \qquad (4)$$

By employing Eqs. (2)-(4) we obtain good fits to the experimental data with $\lambda_{\mathrm{m},0} = 0.98$ μm, $C_0 = 2412$ μΩ, and $\xi = 22.5$ μT$^{-1}$ [see black and red solid lines in Figs. 3(b)-(d)]. Remarkably, the $\lambda_{\mathrm{m},0}$ and $\xi$ values are in agreement with the larger magnon damping expected for TmIG compared to YIG, as $\lambda_{\mathrm{m},0} = 1.86$ μm and $\xi = 0.83$ μT$^{-1}$ were found in YIG films of similar thickness [48].

Finally, we also want to point out that in the large field regime the diffusion constant might be dependent on the magnetic field, $D_\mathrm{m}(B)$, and that $\tau(B)$ may differ from the $(1+\xi B)^{-1}$ dependence assumed above. This is because our relations are derived from a simplified Gilbert damping model that accounts for a linear with field increase of the damping, which is valid for the low frequency magnons, and assumed a constant $D_\mathrm{m}$. Moreover, the spin mixing conductance at the Pt/TmIG interface, which we considered to be constant, could also influence the field dependence of $C$ in the high field range. Note that any of these deviations in the field dependences might also be the reason for the apparent increase of $\lambda_\mathrm{m}^2/C$ in the range from ~700 to 800 mT [Fig. 3(e)]. Nevertheless, we remark that our simple model [Eqs- (2)-(4)] captures very well the field trend of the experimental data presented in Figs. 3(b)-(d). Future experimental and theoretical works should aim at investigating magnon diffusion in other MIs and its dependence at large fields, as well as address the origin of $\lambda_\mathrm{m}(B)$, $n(B)$, and $D_\mathrm{m}(B)$ from a microscopic perspective.

## C. Nonlocal thermoelectric effects

In this section, we analyze the origin of the second harmonic signals. As shown in the schematic of Fig. 4(a), Joule dissipation at the injector electrode not only results in diffusive magnon currents propagating away from the injector/TmIG interface, but also in the build-up of thermal gradients at the detectors. As we will show, both thermal diffusive magnons and nonlocal thermal gradients contribute to the nonlocal second harmonic signal $V_\mathrm{NL}^{2\omega}$, each having a distinct angular, distance, and current dependence.

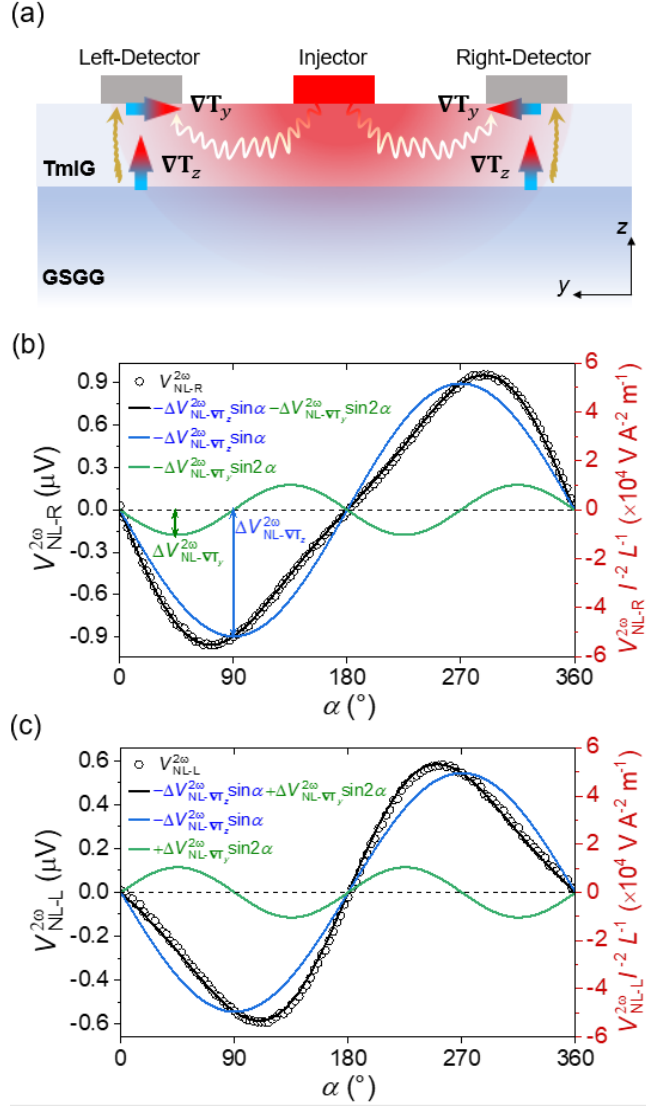

FIG. 4. (a) Cross section schematic of the thermal gradients induced in the nonlocal devices. Magnon diffusion from the injector and from the GSGG/TmIG interface underneath the Pt detectors are schematically represented by wavy arrows. $\nabla T_z$ and $\nabla T_y$ indicate thermal gradients along the $z$ and $y$ directions, respectively. Note that the sketch is not to scale. (b), (c) In-plane angle scans of $V_{NL}^{2\omega}$ measured in detector electrodes located at the right- and left-hand side of the injector, respectively (see Fig. 1(a) for the measurement scheme). Data taken in a device with $d = 1.2$ μm, $I = 310$ μA, and $B = 500$ mT. The scale at the right-hand side presents the voltage data normalized by the length of the detector ($L_R = 185$ μm, $L_L = 125$ μm) and the square of the current. The black lines are fits of the experimental data to a $A \sin \alpha + B \sin 2\alpha$ function. The $\sin \alpha$ contribution is the same for both detectors (blue curves), but the $\sin 2\alpha$ has same amplitude but opposite sign between right and left detectors (green curves). These contributions are referred in the text as $V_{NL-\nabla T_z}^{2\omega}$ and $V_{NL-\nabla T_y}^{2\omega}$, respectively. See (b) for the definition of the amplitudes and the text for details regarding the relative sign of the signals.

Figs. 4(b) and 4(c) demonstrate that for an in-plane angular scan measurement, the $\sin\alpha$ contribution to $V_{\text{NL}}^{2\omega}$ has the same sign and amplitude (see blue curves) for detector stripes located at either side of the injector by the same distance. This is because the $\sin\alpha$ signal is linear with the vertical thermal gradient $\mathbf{\nabla T}_z$ induced at the Pt-injector/MI interface, which generate diffusive magnon currents as well as $\mathbf{\nabla T}_z$ gradients underneath the detectors, the latter resulting in nonlocal SSE-like signals [14] [18]. Therefore, the $\mathbf{\nabla T}_z$ contribution to $V_{\text{NL}}^{2\omega}$ is symmetric for detectors located at either side of the injector [see Fig. 4(a)]. The $\sin2\alpha$ contribution, however, has the same amplitude but opposite sign for Pt stripes located at opposite sides of the injector [see Figs. 4(b) and 4(c), green curves], which strongly indicates that this signal is caused by the in-plane thermal gradient $\mathbf{\nabla T}_y$ present at the detector. In addition to the linear dependence with the thermal gradients, note that the second harmonic voltages are also linear with the SHE at the detector, thus defining the sign of the signals. In the following, we analyze the distance-dependence of the amplitude of these two contributions to $V_{\text{NL}}^{2\omega}$ and determine whether they originate from magnon or heat diffusion as well as whether the thermoelectric signals are of spin or magnetic nature. Last but not least, we also investigate a third thermoelectric contribution, which appears as an anomalous Hall-like response at the detector when sweeping the magnetic field perpendicular to the film plane.

**C.1. Nonlocal spin Seebeck effect**

The origin of the $\sin\alpha$ contribution to the second harmonic signal, which we label as $V_{\text{NL}-\mathbf{\nabla T}_z}^{2\omega}$ (blue curves in Figs. 4(b) and 4(c)), can be determined by analyzing the decay of the signal with the injector-detector distance. Fig. 5(a) shows the amplitude $V_{\text{NL}-\mathbf{\nabla T}_z}^{2\omega}$ as function of $d$ and for different magnetic fields. Unlike the case shown in Fig. 3 for $\Delta V_{\text{NL}}^{1\omega}$, $\Delta V_{\text{NL}-\mathbf{\nabla T}_z}^{2\omega}$ is nearly field-independent, suggesting that $V_{\text{NL}-\mathbf{\nabla T}_z}^{2\omega}$ is originating from the nonlocal thermal gradients present nearby the detector rather than by diffusive magnons generated beneath the injector stripe. This interpretation is supported by the decay of the signal with $d$, which follows a $\sim 1/d^2$ dependence (see Fig. 5(b)) as expected for the radial spreading of heat from the injector to the TmIG/GSGG substrate [18] [due to the geometry of the electrodes, heat diffusion only occurs in the $yz$ plane, see schematic of Fig. 4(a)]. Fig. 10(a) in Appendix D shows the vertical thermal gradient $\nabla T_z$ computed at the TmIG/Pt-detector interface for different $d$ positions of the detector, showing that $\nabla T_z$ indeed decays as $1/d^2$. Only for the shorter $d$ values (1.0 and 1.2 μm), the data slightly deviate from the $1/d^2$ dependence [dashed line in Fig. 5(b)], indicating that for these distances $V_{\text{NL}-\mathbf{\nabla T}_z}^{2\omega}$ might start being influenced by the detection of diffusive magnons as well.

The $1/d^2$ dependence has also been observed in nonlocal Pt devices patterned on YIG films (micrometer-range thick) when $d$ is larger than 3-to-5 times $\lambda_\text{m}$ [14] [18]. This crossover from the exponential to the $1/d^2$ regimes occurs when the thermal diffusion dominates over the diffusive magnon transport, as the latter decays much faster than the heat does (exponential vs radial decay). As a result, the nonlocal thermal gradient leads to a build-up magnon accumulation and diffusion from the MI/dielectric substrate interface. At the detector position, the vertical magnon diffusion results in a SSE-like signal and is thus commonly identified as *nonlocal SSE*. Given that in TmIG we could only explore devices with a minimal distance of 1 μm [~3 times $\lambda_\text{m}$ according to first harmonic measurements, see Figs. 2(b) and 3(c)], it is reasonable that a dominant ~$1/d^2$ dependence is observed in our measurements, and expect that the thermal diffusive magnon transport dominates $V^{2\omega}_{\text{NL}-\boldsymbol{\nabla}\text{T}_z}$ for devices having shorter injector-detector distances.

We finally point out that if only a limited range of injector-detector distances is explored in materials with short $\lambda_\text{m}$ [49] [50], the exponential regime and the $1/d^2$ regime might be confused with each other. This calls for the need to investigate both the first and second harmonic signals to characterize diffusive magnon transport in MIs, as well as to explore $V^{2\omega}_{\text{NL}-\boldsymbol{\nabla}\text{T}_z}$ for a wide range of $d$ values to unambiguously identify the exponential regime associated to thermally-driven magnon currents.

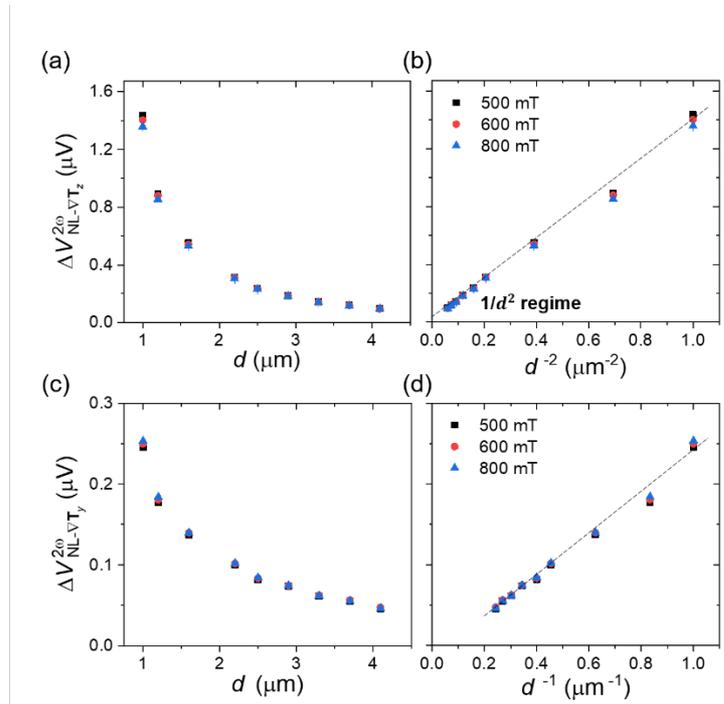

FIG. 5. (a) Amplitude of $V^{2\omega}_{\text{NL}-\boldsymbol{\nabla}\text{T}_z}$ [blue curves in Figs. 4(b) and 4(c)] as function of the injector-detector distance and for different applied fields. The driving current is $I = 310$ μA. (b) Same data in (a) represented as $1/d^2$, demonstrating that $V^{2\omega}_{\text{NL}-\boldsymbol{\nabla}\text{T}_z}$ decays geometrically with $d$ (radially from the injector in the plane perpendicular to the current line), only exhibiting a slight deviation for the shortest injector-detector distances investigated ($d = 1.2$ and $1.0$ μm). The dashed line is a linear fit

to the data with $d \geq 1.4$ μm and extrapolated to large $1/d^2$ values. (c) Same as (a), but for the amplitude of $V^{2\omega}_{\text{NL}-\nabla \mathbf{T}_y}$ [green curves in Figs. 4(b) and 4(c)]. (d) $\Delta V^{2\omega}_{\text{NL}-\nabla \mathbf{T}_y}$ as function of $1/d$. The dashed line is a linear fit to the data. The data in (b) and (d) correlate with the decay with $d$ of the thermal gradients $|\nabla T_z|$ and $|\nabla T_y|$, respectively, at the detector [see Fig. 10(b)]. The length of the detector is $L = 185$ μm. The error in the data points is smaller than the size of the points.

**C.2. Planar/spin Nernst effect**

The sin2$\alpha$ thermoelectric contribution, which we label as $V^{2\omega}_{\text{NL}-\nabla \mathbf{T}_y}$, reverses sign when inverting the position of the detector stripes (see green curves in Figs. 4(b) and 4(c)), indicating that it arises from the in-plane thermal gradient $\nabla T_y$ at the detector. This interpretation is supported by the fact that the amplitude of this signal does not depend on the magnetic field and decays as $1/d$ (see Figs. 5(c) and 5(d)), in agreement with the expected decay of $|\nabla T_y|$ with $d$ (see Fig. 10(b) in Appendix D).

The sin $2\alpha$ dependence is consistent with the expected modulation by **M** of TmIG of the thermally-driven spin accumulation at the detector [22]. The physical mechanism is as follows. $\nabla T_y$ in Pt produces a transverse spin current by the SNE, which is further converted into a transverse charge current due to the inverse SHE. As a result, a thermally-driven longitudinal voltage along $\nabla \mathbf{T}_y$ is generated, an effect that is similar to the conventional Seebeck effect, but fundamentally different as it is driven by the spin-orbit coupling of the HM layer. Unlike the Seebeck effect, the heat-to-spin conversion induced by the SNE results in spin currents that can be modulated via the control of the spin transmission across the TmIG/Pt interface with **M**. The control of the spin transmission leads to a modulation of the longitudinal thermoelectric signal ($\parallel \nabla \mathbf{T}_y$) as well as in a build-up transverse voltage, i.e., along the detector stripe, the latter having a sin $2\alpha$ angular dependence consistent with the experiment [green curves in Figs. 4(b) and 4(c)]. This effect can be regarded as the thermally-driven counterpart of the transverse SMR.

We note that a thermoelectric signal with a sin $2\alpha$ dependence is also expected in metallic magnets in the presence of a transverse thermal gradient, an effect known as PNE [21]. Thus, the $V^{2\omega}_{\text{NL}-\nabla \mathbf{T}_y}$ signal observed in the experiment could also be explained by the PNE at the detector if the Pt atoms at the interface with TmIG become magnetic due to the magnetic proximity effect (MPE) [51]. We investigated this possibility by performing local SMR measurements. The experiments revealed an anisotropic magnetoresistance (AMR) signal of about 10% of that of the SMR amplitude (see Appendix C), thus indicating that in our devices Pt exhibits a non-negligible MPE.

Unlike TmIG/Pt, in bulk YIG/Pt [8] [11] [12] [13] [14] [29] [53] as well as in other MPE-free MI/HM nonlocal magnon transport devices [32] [33] [49] [50], $V^{2\omega}_{\text{NL}-\nabla T_z}$ was found to dominate the second harmonic signal. Experiments in ultrathin YIG(15nm)/Pt films, however, revealed a significant $V^{2\omega}_{\text{NL}-\nabla T_y}$ contribution, which we attribute to the reduction of $V^{2\omega}_{\text{NL}-\nabla T_z}$ relative to $V^{2\omega}_{\text{NL}-\nabla T_y}$ due to the decrease of the magnon diffusion length in YIG thin films [40]. Alternatively, in materials with longer $\lambda_\text{m}$ values, $V^{2\omega}_{\text{NL}-\nabla T_y}$ should become relevant at sufficiently long injector-detector distances, but the decay of both signals with $d$ makes it more difficult to be detected in the experiments. According to Ref. [22] and our estimates of the induced thermal gradients at the detector (Appendix D), we find that $\Delta V^{2\omega}_{\text{NL}-\nabla T_y}$ in our TmIG/Pt devices is about 50% larger than the one expected by the SNE. We thus conclude that both the SNE and the PNE, the latter driven by the MPE, contribute to $\Delta V^{2\omega}_{\text{NL}-\nabla T_y}$, and anticipate that thermoelectric signals driven by the SNE or the PNE will be more easily detected in materials with short magnon diffusion lengths.

Finally, we point out that for nonlocal second harmonic signals of purely thermoelectric origin (i.e., for $d \gg \lambda_\text{m}$, see Section C1), the contribution of $V^{2\omega}_{\text{NL}-\nabla T_y}$ relative to $\Delta V^{2\omega}_{\text{NL}-\nabla T_z}$ increases when increasing injector-detector distance. This is because $\Delta V^{2\omega}_{\text{NL}-\nabla T_z}$ decays as $1/d^2$, whereas $\Delta V^{2\omega}_{\text{NL}-\nabla T_y}$ decays as $1/d$. Therefore, the ratio $\Delta V^{2\omega}_{\text{NL}-\nabla T_y}/\Delta V^{2\omega}_{\text{NL}-\nabla T_z}$ will increase linearly with $d$, making the contribution of $V^{2\omega}_{\text{NL}-\nabla T_y}$ on $V^{2\omega}_{\text{NL}}$ more pronounced as $d$ increases. In fact, in TmIG/Pt devices, it has been shown that for $d = 20$ μm, $\Delta V^{2\omega}_{\text{NL}-\nabla T_y}$ is already larger than $\Delta V^{2\omega}_{\text{NL}-\nabla T_z}$, and that for $d \sim 50$ μm, $\Delta V^{2\omega}_{\text{NL}-\nabla T_z}$ is negligible against $\Delta V^{2\omega}_{\text{NL}-\nabla T_y}$ [24].

### C.3. Thermoelectric anomalous Hall-like signal

In the following, we will show that a third thermoelectric contribution, which arises from the in-plane thermal gradient $\nabla \mathbf{T}_y$ and the MPE at the Pt detector, is detected in the nonlocal second harmonic response when the external magnetic field is swept perpendicular to the film.

Figure 6(a) shows a representative measurement of the nonlocal second harmonic voltage for **B** swept along the out-of-plane direction (see inset schematics). The field trace reveals a thermoelectric anomalous Hall-like (TEAH) voltage $\Delta V^{2\omega}_{\text{NL}-\text{TEAH}}$ [see Fig. 6(a)] that correlates with the orientation of the magnetization of the TmIG layer [$\mathbf{M} = (0,0,M_z)$ points out of the plane and switches at the field values at which $V^{2\omega}_{\text{NL}-\text{TEAH}}$ exhibits voltage jumps; see Appendix B]. A reversal of the position of the detector results in a reversal of $V^{2\omega}_{\text{NL}-\text{TEAH}}(B_z)$ [see Fig. 6(b)], indicating that this effect is linear with $\nabla \mathbf{T}_y$. In metallic ferromagnets, such signal response with $B_z$ is attributed to the anomalous Nernst effect (ANE). This scenario, however, changes with MIs as no electric conduction through the ferromagnet takes place. Similar TEAH signals have also recently been observed in MI/HM heterostructures and attributed to a different physical mechanism, a *thermal spin drag* (TSD) [24] which originates from the combined action of $\nabla \mathbf{T}_z$ and $\nabla \mathbf{T}_y$ at the detector. According to the TSD, the out-of-plane thermal gradient $\nabla \mathbf{T}_z$ at the MI/HM interface results in a spin current propagating along the z-direction with the spin polarization pointing parallel to **M**. The in-plane gradient $\nabla \mathbf{T}_y$ in the HM then drags the spin current along its direction, which is detected as a thermoelectric voltage along the HM stripe due to the subsequent spin-to-charge conversion (note that the inverse SHE voltage is perpendicular to the spin polarization and the spin flow). The TSD is therefore a second order thermoelectric effect, and as such should have a current and distance dependence $\propto \nabla T_z \cdot \nabla T_y$. Given that any linear thermoelectric effect induced by either $\nabla T_z$ or $\nabla T_y$, such as $\Delta V^{2\omega}_{\text{NL}-\nabla\mathbf{T}_z}$ and $\Delta V^{2\omega}_{\text{NL}-\nabla\mathbf{T}_y}$, is $\propto I^2$ (see Appendix E), and that $\nabla T_z$ and $\nabla T_y$ decay as $1/d^2$ and $1/d$ (Fig. 10), respectively, we expect the TSD effect to be proportional to $I^4$ and decay as $1/d^3$.

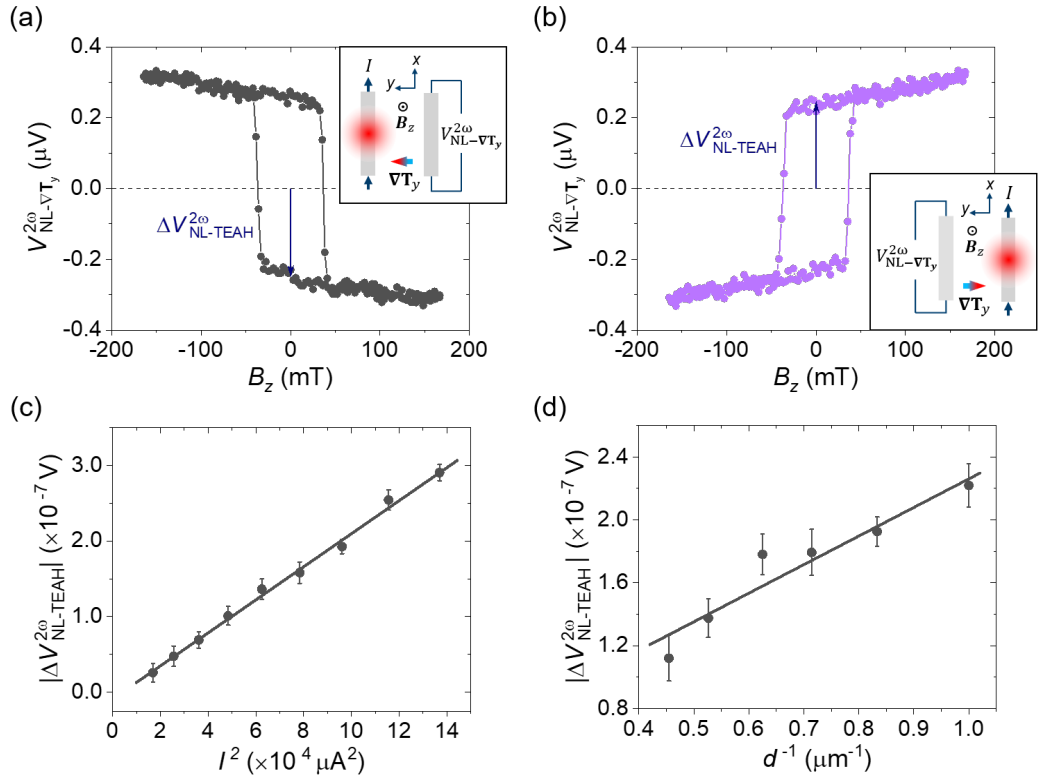

FIG. 6. (a), (b) Nonlocal second harmonic voltage measurements performed at detector electrodes located at the righ- and left-hand side of the injector, respectively, while sweeping the magnetic field perpendicular to the plane. See inset schematics for the measurement configuration. The amplitude of the injected sinusoidal current is $I = 450$ μA, the injector-detector distance $d = 0.5$ μm, and $L_R = L_L = 125$ μm. The linear contribution with $B_z$ is due to the ordinary Nernst effect in Pt, which is linear
16

with $\nabla T_y$ as seen for the sign change between (a) and (b). Additionally, a thermoelectric anomalous Hall-like (TEAH) signal linear with the out-of-plane **M** of TmIG and $\nabla \mathbf{T}_y$ is observed. The amplitude and sign of this signal is indicated as $\Delta V_{\text{NL-TEAH}}^{2\omega}$. A device dependent voltage offset is subtracted for clarity in (a) and (b). (c), (d) Current and distance dependence of $|\Delta V_{\text{NL-TEAH}}^{2\omega}|$ demonstrating that it follows a $I^2$ and $1/d$ dependence, respectively.

Figures 6(c) and 6(d) show that $|\Delta V_{\text{NL-TEAH}}^{2\omega}|$ follows a $\sim I^2$ and a $\sim 1/d$ dependence, thus indicating that only $\nabla T_y$ contributes to the TEAH signal. We thus first consider whether $\Delta V_{\text{NL-TEAH}}^{2\omega}$ is originated by the SNE in Pt and $M_z$ of TmIG through the imaginary component of the spin mixing conductance $G_i$. This physical scenario is equivalent to the thermal counterpart of the anomalous Hall-like SMR signal due to $G_i$. Therefore, in a MPE-free system, the ratio $\frac{\Delta V_{\text{NL-TEAH}}^{2\omega}}{\Delta V_{\text{NL}-\nabla T_y}^{2\omega}}$ driven by the SNE should be comparable to the AHE-like/longitudinal SMR ratio. In our samples, however, we found the latter to be $< 0.02$ (computed from the data in Fig. 8) while $\frac{\Delta V_{\text{NL-TEAH}}^{2\omega}}{\Delta V_{\text{NL}-\nabla T_y}^{2\omega}} \sim 0.57$, indicating that $\Delta V_{\text{NL-TEAH}}^{2\omega}$ cannot be driven solely by the SNE. Moreover, if only the SNE contribution to $\Delta V_{\text{NL}-\nabla T_y}^{2\omega}$ is considered, the ratio $\frac{\Delta V_{\text{NL-TEAH}}^{2\omega}}{\Delta V_{\text{NL}-\nabla T_y}^{2\omega}}$ exceeds 1. This is contrast to measurements in MPE-free samples, where the AHE-like/longitudinal SMR ratio typically falls below 0.1 [54] [55]. As described in Section C.2 and Appendix C, Pt in proximity with TmIG becomes partially magnetic. In this case, and for **M** pointing out-of-the plane, an in-plane thermal gradient results in a $V_{\text{NL-TEAH}}^{2\omega}$ voltage due to the ANE in Pt [23]. We thus alternatively attribute the origin of the abnormally large $V_{\text{NL-TEAH}}^{2\omega}$ response in our sample to the anomalous Nernst effect driven by the MPE and $\nabla \mathbf{T}_y$ at the detector. Remarkably, the ratio ANE/PNE in ferromagnetic metals such as permalloy is similar to the one found in our system [56], further supporting the idea that $\Delta V_{\text{NL-TEAH}}^{2\omega}$ in TmIG/Pt has magnetic origin in contrast to the spin origin proposed by the TSD.

We thus conclude that $V_{\text{NL-TEAH}}^{2\omega}$ in our devices is mainly driven by the MPE and $\nabla \mathbf{T}_y$ at the Pt detector. We also remark that analyzing the current and distance dependence of $V_{\text{NL-TEAH}}^{2\omega}$ is a powerful tool for distinguishing between MPE-induced TEAH signals from those originated by the TSD effect. We also anticipate that the TSD effect would better be detected in the fourth harmonic response because the first order (second harmonic) thermoelectric signals such as the ones originated by the MPE or $G_i$ will be filtered out. We analyzed the fourth harmonic nonlocal Hall voltage in our samples but could not identify any signal within the noise level. We expect that the methodology described here will guide future works aiming at analyzing the origin of thermally-driven anomalous Hall-like signals in MI/HM bilayers.

## IV. CONCLUSIONS

We have investigated magnon transport and nonlocal thermoelectric effects in perpendicularly magnetized ultrathin TmIG/Pt nonlocal devices. We have demonstrated the electrical detection of diffusive magnon spin signals to distances exceeding 1 μm and showed that Joule heating leads to a number of nonlocal thermoelectric effects. Remarkably, we have shown that by performing angle, current, and distance dependent nonlocal and local harmonic transport measurements it is possible to distinguish between spin- and thermally-driven magnon signals as well as to identify the origin of diverse thermoelectric voltages in MI/HM devices. In TmIG/Pt, we found that the out-of-plane thermal gradient at the TmIG/Pt-detector interface results in a nonlocal SSE, whereas the in-plane thermal gradient results in SNE, PNE, and ANE thermoelectric voltages. While the SNE is intrinsic of Pt, the PNE and the ANE emerge from the MPE at the TmIG/Pt interface. We also showed that nonlocal thermoelectric effects are more easy to be detected in devices exhibiting short magnon diffusion lengths (sub-μm range), and that the SNE, PNE, and ANE will eventually dominate over the nonlocal SSE in the long-distance regime. We finally discussed how to distinguish between the TSD and the ANE and demonstrated that in our devices the TSD is negligible. It is worth noting that the thermoelectric effects investigated here are not exclusive to magnon transport devices but expected to be present in any MI/HM heterostructure whenever thermal gradients are present.

We furthermore showed that the magnon diffusion length $\lambda_\mathrm{m}$ and the distance-independent scaling prefactor of the nonlocal signal $C$ in TmIG decrease with the magnetic field, which we attribute to the suppression of $\tau$ and $n$ with increasing magnetic field. We derived simple relations for $\tau(B)$, $n(B)$, $\lambda_\mathrm{m}(B)$, and $C(B)$ based on a linear increase of the Gilbert damping with field and comparison with the experimental data, and showed that they can nicely reproduce $V_\mathrm{nl}(B,d)$ in the field range explored. Future work should aim at addressing the microscopic origin of the field dependence of the magnon transport, as well as determine whether other parameters such as $D_\mathrm{m}$ also contribute to the magnetic field dependence.

Finally, our work validates sputter-growth as a suitable alternative to pulsed laser deposition to fabricate high-quality thin film oxides and oxide/metal bilayers for spintronic applications. The possibility of combining magnon transport with other device functionalities such as current-induced switching and domain wall motion in perpendicularly magnetized layers coupled to HMs open prospects for developing novel device concepts.

**APPENDIX A: CRYSTALLOGRAPHIC AND TOPOGRAPHIC CHARACTERIZATION OF THE TMIG FILMS**

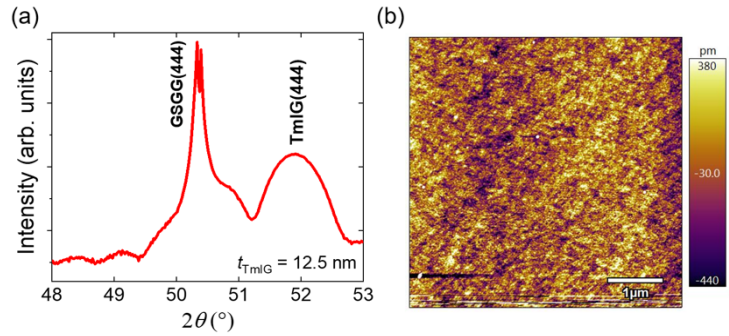

FIG. 7. (a) Structural characterization of a representative TmIG (12.5 nm) thin film. The $2\theta - \omega$ X-ray diffraction scan around the (444) peak of TmIG reveals a Laue diffraction peak and Laue oscillations characteristic of fully strained epitaxial films [4] [27]. (b) Surface topographic characterization of a representative TmIG (30nm) film via atomic force microscopy (AFM). The root-mean-square roughness is less than 1 nm over a $5 \times 5$ µm² surface area. Films with a Pt (4nm) top layer exhibit the same roughness.

**APPENDIX B: ELECTRIC CHARACTERIZATION OF THE MAGNETIC ANISOTROPY AND THE COERCIVE FIELD OF TMIG**

The magnetic anisotropy of TmIG films can be characterized by SMR measurements [4] [27] [28] [29]. Fig. 8(a) shows an optical image of a Pt Hall bar patterned on the TmIG (15nm) film. The electric wiring employed in the local transport measurements is indicated. Fig. 8(b) shows a transverse magnetoresistance measurement while sweeping the out-of-plane field $B_z$. The measurement reveals a clear squared-shaped anomalous Hall-like response, demonstrating that the films exhibit robust perpendicular magnetic anisotropy with a coercive field $H_c \sim 30$ mT. Fig. 8(c) shows the transverse magnetoresistance measured for **B** applied in the plane of the film and at an angle $\alpha = 45$ and 135° from the current line (red and blue solid dots, respectively). The gradual change of resistance from $B = 0$ to $\sim 400$ mT is due to the gradual tilt of **M** of TmIG towards **B**, thus indicating that the anisotropy field of TmIG is $\sim 400$ mT. Note that the signals at 45 and 135° have same amplitude but opposite sign due to the $\sin 2\alpha$ symmetry of the transverse SMR.

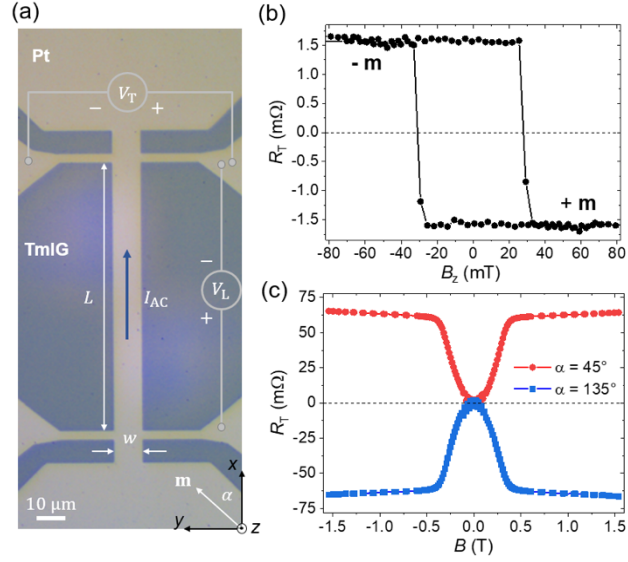

FIG. 8. (a) Optical image of a Pt Hall bar patterned on the 15-nm-thick TmIG film investigated in this work. The electric wiring employed in the measurements, the coordinate system, and the direction of field rotation are indicated. (b) Transverse magnetoresistance measurement $R_T = V_T/I$ performed while sweeping the out-of-plane field $B_z$. The coercive field is $H_c \sim 30 \text{ mT}$. (c) Transverse magnetoresistance measurements performed with a magnetic field applied in the plane of the film and at an angle $\alpha = 45°$ (red solid dots) and $135°$ (blue solid dots) from the current line. See panel (a) for the definition of $\alpha$. The graph contains both the trace and retrace curves. From these measurements, we determine the anisotropy field to be $\sim 400 \text{ mT}$.

**APPENDIX C: LONGITUDINAL SMR MEASUREMENTS – MPE IN PT ON TMIG**

The MPE in Pt can be identified from longitudinal SMR measurements [29]. Fig. 9 shows angular dependent longitudinal SMR measurements performed along the three main axes of the device. According to SMR, the magnetoresistance modulation along the $xy$ and $yz$ planes should be identical, and no magnetoresistance should be visible in the $xz$ plane [29] [57] (see sketches in Fig. 9a for the definition of the rotation planes). However, a magnetoresistance amplitude $\sim 10\%$ of that observed in the $xy$ plane is detected in the $xz$ plane, which indicates the presence of a finite AMR. We rule out that this signal arises from ordinary magnetoresistance in Pt films as the resistivity in our devices is $\sim 45$ μΩ·cm (a resistivity of about 20 μΩ·cm would be necessary [58] [59]). The presence of AMR thus indicates that the Pt atoms in proximity with TmIG become magnetic. The presence of MPE at MI/HM interfaces has been a topic of intense debate in the last decade [3] [28] [51] [55] [60] [61] [62] [63], now accepted to be strongly dependent on the materials involved and the deposition conditions. In TmIG/Pt, previous magnetotransport studies did not report clear evidence of MPE [4] [5] [24] [25] [26] [27], but a surprisingly large spin mixing conductance as well as a large induced magnetic moment per Pt atom at the interface with TmIG were estimated from magnetometry measurements [4]. It is therefore plausible that in our devices the MPE at the TmIG/Pt interface induces a magnetic moment in Pt.

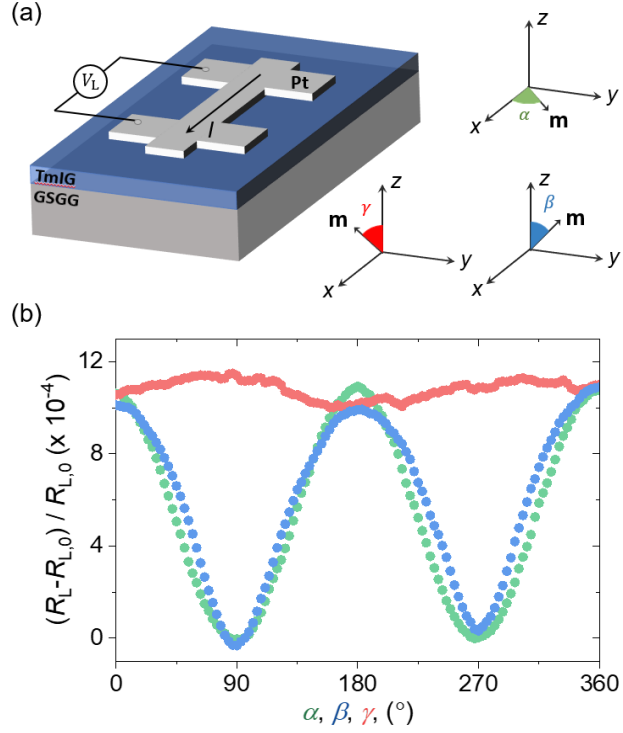

FIG. 9. (a), (b) Longitudinal spin Hall magnetoresistance performed along the three main axes of the device. See sketch in (a) for the definition of the angles, the color-code of the measurement plane, and the coordinate system. An optical image of the device and the measurement scheme employed is shown in Fig. 8(a). The SMR amplitude is defined as $\Delta R_L = \frac{R_L - R_{L,0}}{R_{L,0}}$, where $R_L = V_L/I$ and $R_{L,0} = R_L(\mathbf{B}||\mathbf{y})$. The amplitude of the magnetoresistance in the $xz$ plane is ~10% of that measured in the $xy$ plane. We ruled out that this signal arises from a misalignment of the sample with the field. We thus conclude that TmIG induces a magnetic moment in Pt due to proximity at the interface.

**APPENDIX D: COMPUTATION OF $\nabla T_z$ AND $\nabla T_x$ AT THE DETECTOR**

By employing a finite element analysis simulation (MATLAB), we compute the thermal gradients $|\nabla T_z|$ and $|\nabla T_y|$ at the detector position. Because of the symmetry of the system, only thermal diffusion in the $yz$ plane is taken into account. For the simulation, we considered the current injector as a heat source and compute the thermal gradients $\nabla \mathbf{T}_z$ and $\nabla \mathbf{T}_y$ at different positions across the heterostructure [Fig. 4(a) shows a schematic of the device layout; note that the thickness of TmIG is not represented at scale]. In the simulation, we considered the dimensions of the electrodes as the ones employed in the experiment, and the thermal conductivities of TmIG/GSGG and Pt to be 8 W/mK and 26 W/mK, respectively [23]. The thickness of the GSGG substrate was considered semi-infinite as it is larger than the lateral dimensions of the experiment. We consider the current line to dissipate

heat at an average rate of 1.5 mW, which corresponds to the power dissipation of an ac current of amplitude $I = 310$ μA at the injector electrode, a characteristic value employed in the experiments. $\nabla T_z$ and $\nabla T_y$ are computed as the average of the out-of-plane and in-plane thermal gradient values in TmIG and Pt, respectively, at the region covered by the detector electrode. Figs. 10(a) and 10(b) show the computed $|\nabla T_z|$ and $|\nabla T_y|$, showing that they decay with the injector-detector distance as $1/d^2$ and $1/d$, respectively.

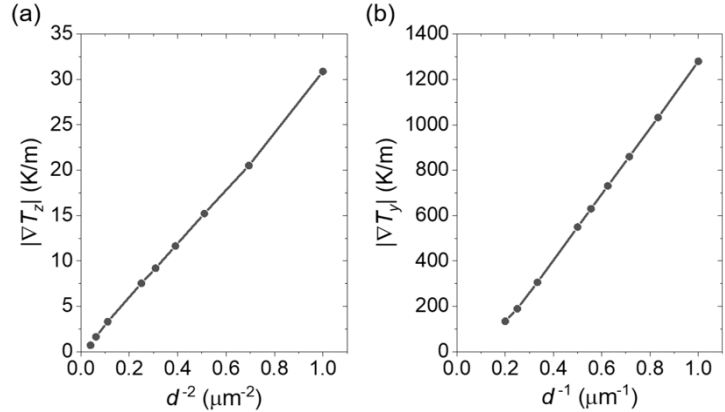

FIG. 10. (a) Average out-of-plane thermal gradient $|\nabla T_z|$ underneath the detector as function of $1/d^2$. (b) Average in-plane thermal gradient $|\nabla T_y|$ at the detector as function of $1/d$. Here we consider a heat power dissipation of 1.5 mW at the injector. See text for more details regarding the simulations.

**APPENDIX E: CURRENT DEPENDENCE OF $V_{NL}^{1\omega}$, $V_{NL-\nabla T_z}^{2\omega}$, AND $V_{NL-\nabla T_y}^{2\omega}$**

The first and second harmonic nonlocal voltage response are odd and even with the ac current excitation, respectively [8]. Therefore, at the lowest order, they are linear and quadratic with $I$. Fig. 11(a) shows the current dependence of $\Delta V_{NL}^{1\omega}$ demonstrating that it is linear with $I$. Figs. 11(b) and 11(c) show the current dependence of $\Delta V_{NL-\nabla T_z}^{2\omega}$ and $\Delta V_{NL-\nabla T_y}^{2\omega}$ for devices having different injector-detector distances, demonstrating that the signals are proportional to $I^2$ as expected.

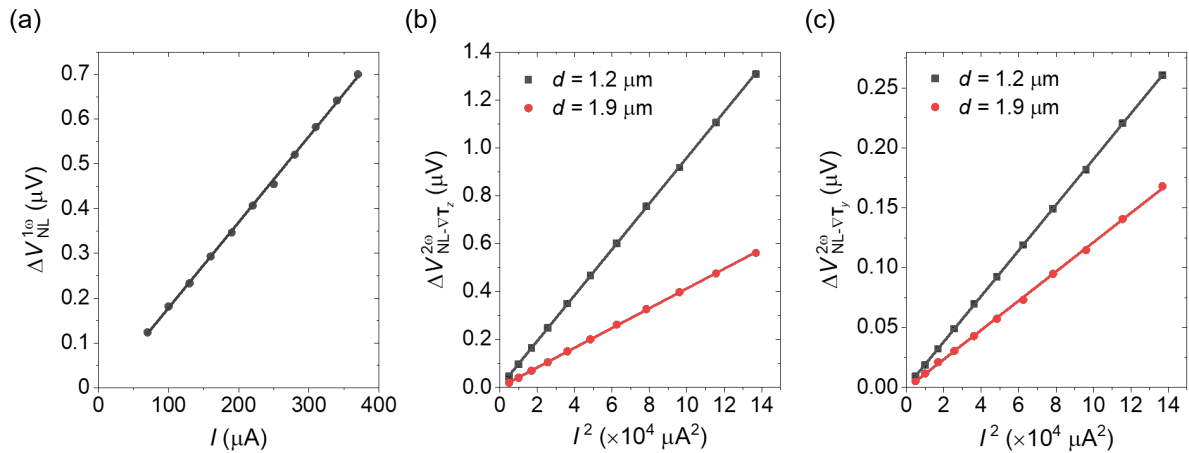

FIG. 11. (a) First harmonic nonlocal voltage measured as function of the ac current $I$. Data taken in a device with $d = 1.2$ μm and $L = 185$ μm. (b), (c) $\Delta V_{\text{NL}-\nabla T_z}^{2\omega}$ and $\Delta V_{\text{NL}-\nabla T_y}^{2\omega}$, respectively, as function of $I^2$ measured in devices with different injector-detector distances and fixed detector length $L = 185$ μm. $B = 500$ mT in all measurements.

**ACKNOWLEDGEMENTS**

This project has received support by the Swiss National Science Foundation (SNSF) via Project Nos. 198642, 20020-172775, and 200021-178825, by the European Research Council through the Advanced Grant 694955-INSEETO, and by the ETH Zürich through the Career Seed Grant SEED-20 19-2. S.V. acknowledges support by the Ministry of Science, Innovation and Universities through the 'Maria de Maeztu' Programme for Units of Excellence in R&D Grant No. CEX2018-000805-M. J. G. acknowledges support from the IDEA scholarship and the Zeno Karl Schindler Foundation.